\documentclass[aps,prd,twocolumn,showpacs,superscriptaddress,nofootinbib]{revtex4-1}
\usepackage[utf8x]{inputenc}
\usepackage{graphicx}  
\usepackage{dcolumn}   
\usepackage{bm}        
\usepackage{amssymb}   
\usepackage{color}
\usepackage{aas_macros}

\newcommand{\msun}{\ensuremath{\mathrm{M}_\odot}}
\newcommand{\sv}[1]{\textcolor{black}{#1}}

\newcommand{\Event}{GW150914}
\newcommand{\Xmas}{GW151226}

\newcommand{\WF}{IMRPhenomPv2}

\newcommand{\si}{\ensuremath{\sim}}
\newcommand{\mc}{\ensuremath{\mathcal{M}}}

\newcommand{\beq}{\begin{equation}}
\newcommand{\eeq}{\end{equation}}

\newcommand{\dg}{\ensuremath{^\circ}}
\newcommand{\paper}{paper}

\begin{document}
\title{Three observational differences for binary black holes detections with second and third generation gravitational-wave detectors}
\author{Salvatore Vitale}
\affiliation{LIGO, Massachusetts Institute of Technology, Cambridge, Massachusetts 02139, USA}

\begin{abstract}

Advanced gravitational-wave observatories, such as LIGO and Virgo, will detect hundreds of gravitational wave signals emitted by binary black holes in the next few years. The collection of detected sources is expected to have certain properties.
It is expected that a selection bias will exist toward higher mass systems, that most events will be oriented with their angular momentum pointing to or away from Earth, and that quiet events will be much more numerous than loud events. 
In this \paper{} we show how all these assumptions are only true for existing detectors and do not have any universality.
Using an network of proposed third-generation gravitational wave detectors, we show how each of these assumptions must be revised and we discuss several consequences on the characterization of the sources.
\end{abstract}
\maketitle

\section{Introduction}

The LIGO observatories have recently detected the gravitational waves (GWs) emitted by the binary black holes (BBHs) \Event{}~\cite{GW150914-DETECTION} and \Xmas{}~\cite{GW151226-DETECTION,O1BBH}, starting the era of gravitational wave astrophysics. Over the next few years, hundreds of similar systems will be detected~\cite{GW150914-RATES,O1BBH}, allowing for studies of formation channels of compact binaries (CBC) and stellar evolution~\cite{GW150914-ASTRO,2016arXiv160808223M,2015arXiv150304307V,2016ApJ...824L...8R,2016MNRAS.458.2634M,2016PhRvD..93h4029R}, tests of general relativity~\cite{GW150914-TESTOFGR,2016arXiv160308955Y,2016PhRvD..94b1101G} and characterization of black holes (BHs) mass and spins~\cite{VitalePrep,2016arXiv160808223M}.

At the same time, R\&D is ongoing to design the next generation of GW detectors, which would add another factor of 10 over the sensitivity of current instruments (which we will refer to as second-generation or 2G). The Einstein Telescope~\cite{2010CQGra..27s4002P,2011CQGra..28i4013H} and the Cosmic Explorer (CE)~\cite{2016arXiv160708697A} are two proposed designs for third-generation (3G) detectors.
Owing to their astonishing sensitivity, 3G detectors would be able to observe BBHs up to redshifts of 10 and above~\cite{Vitale3G}.

In this \paper{} we show how this has deep consequences on the characteristic of detected BBHs.
In particular we show how three facts which are usually assumed as self-evident, are in reality only due to the limited range of 2G detectors and will not be true anymore when 3G detectors come online.

\section{Difference \#1: mass selection bias}\label{Sec.Masses}

It is well-known that the amplitude of GWs emitted by CBCs goes (at the lowest order of the inspiral) as $\mathcal{M}^\frac{5}{6}$,  where \mc{} is the chirp mass defined as $\mathcal{M}\equiv (m_1 m_2)^{3/5}/(m_1 + m_2)^{1/5}$. This makes systems with higher mass easier to detect.
At the same time, the duration of the signal decreases with the total mass (compare e.g. \Event{} and \Xmas~\cite{O1BBH}), which makes very massive systems hard to detect (or undetectable, if so massive that they merge before reaching the lower-frequency side of the band of ground-based detectors).
These two effects work one against the other, resulting in a non-trivial detection efficiency curve as function of the total mass (or chirp mass). In other words, there will be a selection bias for some values of mass. \sv{This selection effect was naturally taken into account by the LIGO and Virgo collaborations when inferring formation rates~\cite{GW150914-RATES,GW150914-RATESSuppl} and mass distributions~\cite{O1BBH} from the events detected in the first science run.}

However it is important to stress that what sets the evolution of the waveform in the detectors' band is the \emph{redshifted} mass, which is larger than the intrinsic (or source-frame) mass by a factor of $(1+z)$~\cite{Maggiore}. This has non-trivial consequences on which systems will be detected more often by 2G and 3G detectors. 

Let us consider a population of BBHs detected by a network of 2 advanced detectors and a network of 2 CE detectors. In both cases we use the geographical coordinates of the two LIGO sites, but our results do not depend on this choice. The noise spectral densities we have used are shown in Ref.~\cite{2016arXiv160708697A}.
We consider BBH sources with intrinsic total mass uniform in the range $[12-200]$~\msun. The lower limit of this range is due to the evidence that stellar-mass BHs might have masses above $\sim5$~\msun~\cite{2011ApJ...741..103F}. The upper limit is somewhat arbitrary, due to the lack of observational evidence for intermediate-mass BHs (IMBH).  The mass ratio is uniform in the range $[0.3-1]$, consistent with the range of validity of the waveform approximant we use (\WF~\cite{Hannam:2013oca,2016PhRvD..93d4007K,2016PhRvD..93d4006H}) and the fact that several astrophysical formation scenarios point to mass ratio distributions larger than 0.5~\cite{2016PhRvD..93h4029R,2016Natur.534..512B}.
One could consider different mass distributions, e.g. power law. The reason why we prefer a uniform mass distribution is that it clearly shows which trends exist, without introducing selection effects from quantities, such as the mass function, which are not yet firmly known. In Appendix~\ref{App.Power}  we discuss how our findings would change in another mass function were chosen.

\begin{figure}[htb]
\includegraphics[width=0.9\columnwidth]{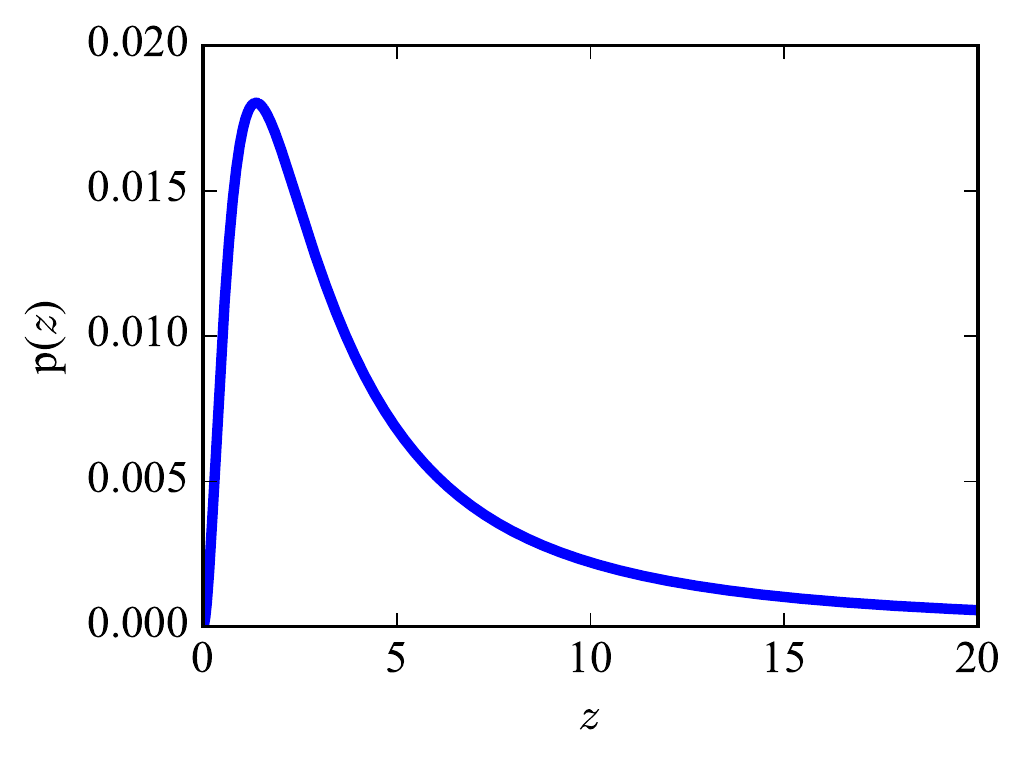}
\caption{The redshift distribution used to generate the simulated events, before the SNR cut.}
\label{Fig.RedshiftPrior}
\end{figure}

The population is generated as follow. We first draw a random set of masses (in the range specified above), spins (uniform in magnitude in the range $[0,1]$, i.e. from non-spinning to maximally spinning,  and orientation), orbital orientation and sky position. The redshift is generated uniform in comoving volume, Fig.~\ref{Fig.RedshiftPrior}, and converted into a luminosity distance assuming a standard $\Lambda$CDM cosmology~\cite{2015arXiv150201589P}.  
This off course assumes that the rate of BBH is flat through cosmic history, which is not strictly true (see e.g. Refs.~\cite{2016Natur.534..512B,2013ApJ...779...72D} for recent estimate). We will come back to this point later.

Only events with network signal-to-noise ratio (SNR) in the range $[10,600]$ are kept. We are assuming that events with SNR below 10 would hard to detect with confidence, while events with SNR over 600 would be rare (not in absolute terms, but when compared to the frequency at which smaller SNRs events are detected, see Fig.~\ref{Fig.SNRs} below). 

\begin{figure}[htb]
\includegraphics[width=0.9\columnwidth]{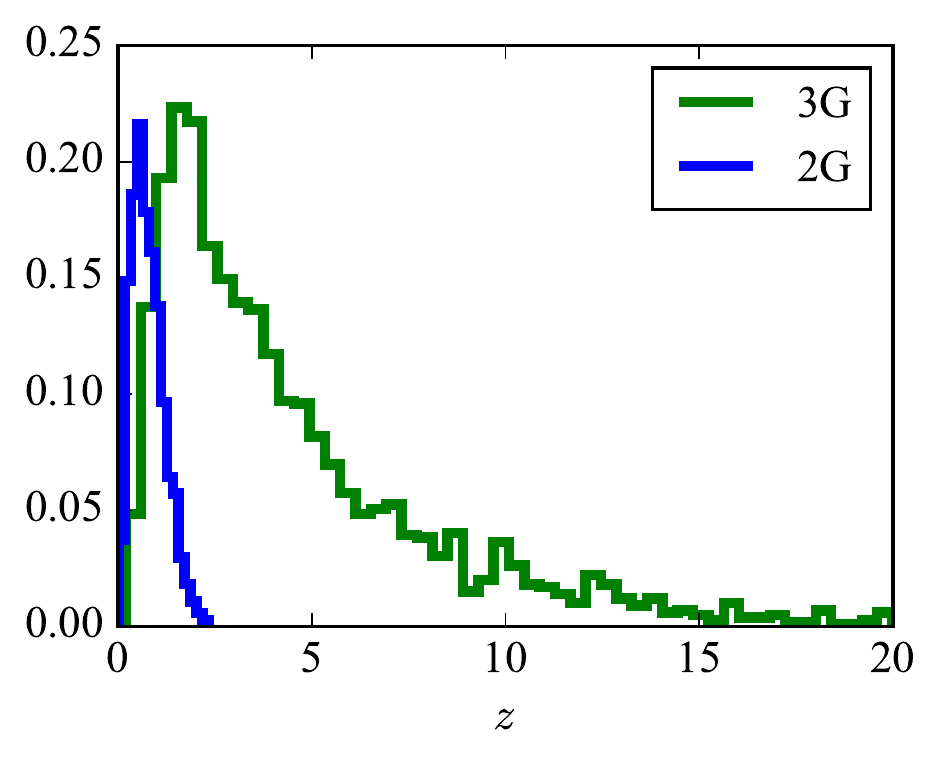}
\caption{The redshift distribution of detectable events with a 2-detector network of advanced detectors at design (2G) or CE-like (3G). Note that the two curves use different y scales to improve clarity.}
\label{Fig.RedshiftComparison}
\end{figure}

In Fig.~\ref{Fig.RedshiftComparison} we report the redshift distribution of the events that survive the SNR cut. As expected, our hypothetical 2G network would be sensitive to BBH (and small IMBH) up to redshift of \si2, while 3G instruments will get something very similar to the prior, Fig.~\ref{Fig.RedshiftPrior}.

\begin{figure}[htb]
\includegraphics[width=0.9\columnwidth]{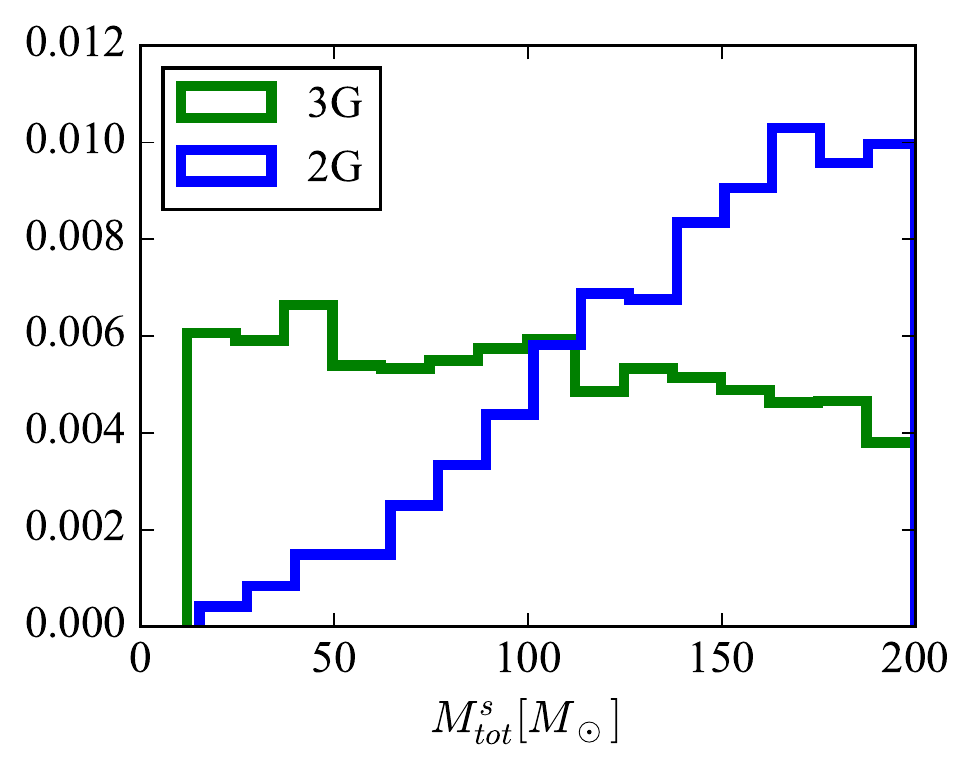}
\caption{The source-frame total mass distribution  of detectable events with a 2 interferometers network of advanced detectors at design (2G) or CE-like (3G).}
\label{Fig.SourceMass}
\end{figure}

Let us now consider the distribution of the \emph{source-frame} BBH total mass, Fig.~\ref{Fig.SourceMass}. A clear difference is apparent between 2G and 3G detectors. While the 2G will on average detect high mass BBH more often than light systems, for 3G the distribution has slightly more support for lower masses, and is similar to the prior we assumed.

For 2G detectors, small mass BBH can only observed nearby, while heavier systems can also be detected farther away, where there is more volume, hence more events. That won't be true anymore for 3G networks, because light systems can also been observed far way, with redshifted masses that will make them equivalent to heavier BBH in the same volume element. Since the bulk of the events will come from the region of universe where there is ``most space" at around $z\sim 2$, what sets the relative rate of detection of BBHs below 200\msun{} in 3G networks will not be the noise of the instrument, but the relative abundance of the systems at those redshifts.

The reversal of the selection bias would be present even if we had used a different mass function. In Appendix~\ref{App.Power} we will see that it would last if we had used a power law mass function that favors stellar-mass BBH. Here we mention that the bias would be even stronger if IMBH were common. In fact, if we had left the high end of the $M^s_{tot}$ distribution go up to $\sim2000$\msun{} we would have found that the rate at which 3G instruments would detect BBH will \emph{decrease} monotonically with the intrinsic total mass. For 2G, instead, it will first go up, till $M^s_{tot}\sim400$\msun, and then go down with a long tail that extends to \si1500\msun. A similar non-monotonic behavior was naturally also present in 1G detectors~\cite{2014PhRvD..89l2003A}.
 
The reason why very heavy sources would be hard to detect for 3G is that at $z\sim2$ they would have a detector-frame total mass a factor of 3 larger, making them coalesce at too low frequencies.

\section{Difference \#2: signal-to-noise ratio}\label{Sec.SNR}

\begin{figure}[htb]
\includegraphics[width=0.9\columnwidth]{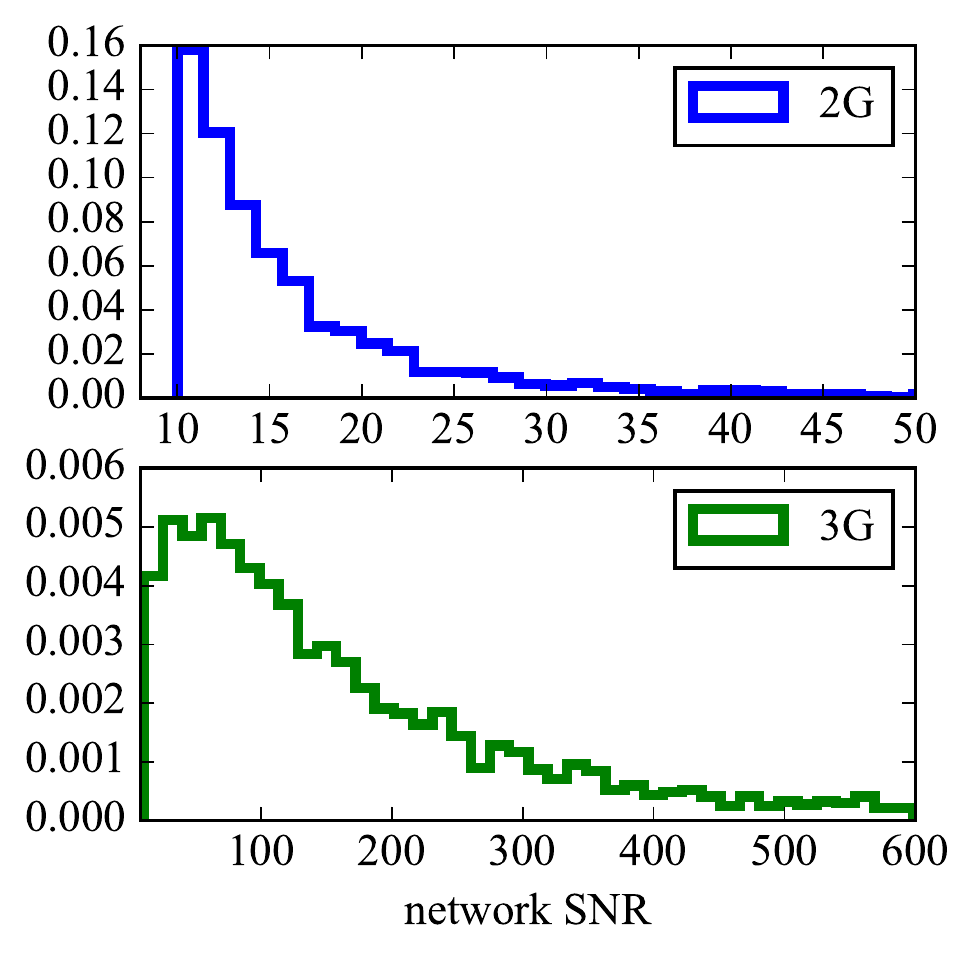}
\caption{The network SNR distribution of the events in Fig.~\ref{Fig.RedshiftComparison}}
\label{Fig.SNRs}
\end{figure}

Let us now consider the distribution of SNR for detected BBH in 3G versus 2G networks.
With  2G detectors it is commonly expected that most sources will be far away, where more volume is available which would lead to detectable events. This, in turn, results on a distribution for the network SNR which peaks at low SNRs, and then goes down as SNR$^{-4}$, Fig.~\ref{Fig.SNRs} top panel. 
This has led some authors to propose using eventual discrepancies between the measured and the expected SNR distribution as a way of testing general relativity~\cite{2016CQGra..33p5004C}.

For 3G instruments the situation is radically different, since they will detect most events at redshifts of a few (where more volume is available) with high SNRs. The peak of the SNR distribution for 3G, thus, is \emph{not} reached for threshold events, but for larger SNRs of \si 70, Fig.~\ref{Fig.SNRs} bottom panel. The ``missing'" low SNR events would have to be at much larger redshifts, where there is less space.
With 3G detectors, thus, it will \emph{not} be the case that the typical BBH source will be at threshold~\footnote{We notice that this difference in the SNR curves would still be present if the extended BBH mass range described at the end of last section or the power law mass function on Appendix~\ref{App.Power} were used.}. Suppression of quiet events was already mentioned by Ref.~\cite{2014arXiv1409.0522C} in the context of a Einstein Telescope single-instrument analysis and for some particular values of masses.

We stress that louder SNR does not automatically imply that parameter estimation will \emph{always} be better for 3G than it is for 2G for systems of similar intrinsic mass. This happens because the masses can be considerably redshifted for 3G, leading to shorter chirps which can compensate for the extra SNR. However, for events at redshift below \si3 a network of three 3G instruments would yield better estimation for both masses and spins~\cite{Vitale3G,VitalePrep}.

\section{Difference \#3: inclination angle}\label{Sec.Iota}

\begin{figure}[htb]
\includegraphics[width=0.9\columnwidth]{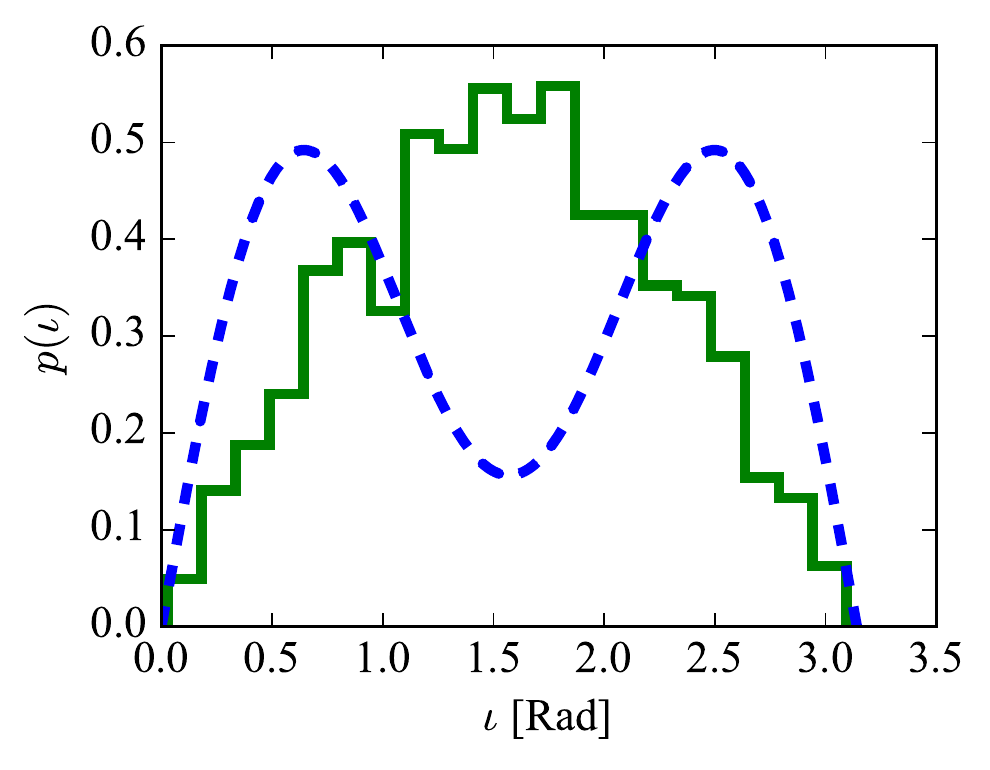}
\caption{The distribution of inclination angles for the events in Fig.~\ref{Fig.RedshiftComparison} (full line) compared with the expected distribution for 2G detectors (dashed line, from~\cite{2011CQGra..28l5023S}) }
\label{Fig.Iota}
\end{figure}

We now consider the inclination angle, i.e. the angle between the orbital angular momentum and the line of sight. 
It is widely accepted~\cite{2011CQGra..28l5023S} that this angle will be close to \si30~degs (or 150~degs) for the typical detected event. 
This happens because two factors are at play: from one side, sources randomly oriented should have a distribution of inclinations that goes like $\sin\iota$; from the other side, since more GW energy goes along the orbital angular momentum~\cite{Maggiore}, sources face-on ($\iota=0$) or face-off ($\iota=\pi$) can be seen farther away, and hence be detected more often.
These two factors result in a bimodal distribution for $\iota$, shown in Fig.~\ref{Fig.Iota} as a dashed line (this is eq. 28 of Ref.~\cite{2011CQGra..28l5023S}~\footnote{We notice that we multiplied eq. 28 of Ref.~\cite{2011CQGra..28l5023S} by $\sim 2$ to properly normalize the probability in the range $[0,\pi]$ }).

After the discussion in the previous sections, it should not surprising that in reality the dashed curve in Fig~\ref{Fig.Iota} is not a universal distribution, but rather due to the limited reach of 2G detectors, and the fact that much more space is available beyond their range.
Indeed, this will be different with 3G detectors. As seen in the discussion about the SNR here above, detections will not be dominated by weak events far away (and hence likely face-on/off), but rather from loud events in the bell of the redshift distribution, Fig.~\ref{Fig.RedshiftPrior}. Since those have random orientation (i.e. $p(\iota)\sim \sin\iota$), the overall distribution will peak at $\pi/2$, Fig.~\ref{Fig.Iota} continuous line.

We can thus expect that the orientation of BBH detected with 3G detectors will roughly be random. This has a few interesting consequences.

The inclination angle has an impact on the uncertainty in the measurement of spin and mass parameters. In fact, eventual spin-induced precession is suppressed in systems within a few tens of degrees from face-on or face-off, whereas visible precessing would break correlations and decrease errors~\cite{PhysRevLett.112.251101,VitalePrep}. Indeed, for both \Event{} and \Xmas{} the posterior distribution for $\iota$ was consistent with face-on/off, and the spins were poorly estimated~\cite{GW150914-PARAMESTIM,O1BBH,GW151226-DETECTION}. The distribution we obtained for 3G instead implies that most events will have large visible precession. 

It has recently been shown by Ref.~\cite{2016PhRvL.117f1102L} how 2G detectors could be able to detect  the Christodoulou GW memory~\cite{1987Natur.327..123B,1991PhRvL..67.1486C}, given enough detections.  Many detections are required because the memory effect is zero for face-on/off sources~\cite{2009ApJ...696L.159F} and near its minimum  for small inclination angles. Conversely, the memory effect is strongest for edge-on systems~\cite{2009ApJ...696L.159F}, making 3G detectors much better suited at measuring it.

Analysis of the ringdown modes of the final BHs formed from the mergers in BBHs is viable way to test the no-hair theorem~\cite{2012PhRvD..85l4056G,2014PhRvD..90f4009M}. This too is a test that will benefit from the distribution of inclinations we found, since the weights of the higher order ringdown modes, needed for the test, are larger for inclination angles close to $\pi/2$~\cite{2009CQGra..26p3001B,2006PhRvD..73b4013B}

\sv{It is known that luminosity distance and orbital inclination angle are correlated in CBC signals, with characteristic V shaped 2-D posterior distributions (e.g. Fig. 3 of Ref.~\cite{PhysRevD.75.062004}). For sources whose orientation is close to edge-on, one can expect better estimation of both these parameters. 3G detectors will thus yield better distance estimates than 2G detectors, which could help associate GW sources to their host galaxies, and estimate cosmological parameters~\cite{2012PhRvD..86d3011D}.}

It has recently been suggested that the kick velocity of the BH created by a BBH coalescence could be measurable with 3G instruments~\cite{2016PhRvL.117a1101G}. The best sources for that measurement are systems oriented face-on or face-off at merger since they will have the largest measurable recoil speed~\cite{2016PhRvL.117a1101G}. 3G detectors would thus typically detect signals with small visible recoil velocity.

Finally, We notice how this could also impact the probability of joint EM and GW detections. If BBH are luminous and their EM radiation is collimated around the orbital angular momentum, similarly to what expected for GRBs in BNS and NSBH (See Ref~\cite{2012ApJ...746...48M} for a review), then with 3G detectors only a small fraction of detected sources would be oriented such that the beam would intersect the Earth. This should only marginally affect BNS with 3G detectors, given to their smaller masses. 

We end this section by noticing that even for 2G detectors, as they get more sensitive in the next few years~\cite{2016LRR....19....1A}, it will be the case that the distribution of inclinations for detected heavy BBH will have more events a $\pi/2$ than the dashed curve of Fig.~\ref{Fig.Iota} suggests, although not fully isotropic.

\section{Conclusions}

In this \paper{} we have described three important properties commonly expected from the population of binary black holes detectable with gravitational-wave detectors and shown how they will not be true anymore when more sensitive detectors come online.
In fact, it is commonly accepted that the distribution of signal-to-noise ratio for detected events will be monotonically decreasing, so that the ``typical'' source will be weak. Relatedly, most events are expected to have inclination angles close to $30$~\dg{} and $150$~\dg. 
We show how both these facts are not universal properties of the sources, but rather depend on both the sources and the instruments used. The next generation of GW detectors, such as the Einstein Telescope and the Cosmic Explorer will detect more BBH sources at large SNRs than at threshold, and most of them with inclination angles close to $90$\dg. We showed how this has positive consequences on the characterization of black holes, making it easy to measure their quasinormal modes, spins and masses. Furthermore, the Christodoulou memory effect will also be more easily measurable, since its SNR is maximum for inclinations of $90$\dg. 
Finally, we showed how the direction of the mass selection bias will be reversed. If the mass function of BBH were flat, 2G detectors would preferentially detect BBH of higher mass (up to \si500\msun{} total, after which the efficiency decreases again), whereas 3G detectors would preferentially detect lower mass systems. 

\sv{We conclude by mentioning that neutron star binaries, owing to their small mass, will not present with any of the effects analyzed in this paper, unless their formation rate density has negligible support for redshifts above \si1, which does not seem the case, based on the redshift distribution of known short GRBs.~\cite{2014ARA&A..52...43B,2016A&A...594A..84G}.}

\section{Acknowledgements}

The author would like to thank Marica Branchesi, Hsin-Yu Chen, Jolien Creighton, Thomas Dent, Reed Essick, Matthew Evans, Davide Gerosa, Daniel Holz, Scott Hughes, Erik Katsavounidis, Nergis Mavalvala, Riccardo Sturani, John Veitch and Mike Zucker for useful discussions and suggestions.
I also thank the anonymous referee of PRD for very helpful comments and suggestions.
The author acknowledges the support of the National Science Foundation and the LIGO Laboratory. LIGO was constructed by the California Institute of Technology and Massachusetts Institute of Technology with funding from the National Science Foundation and operates under cooperative agreement PHY-0757058.
This is LIGO Document P1600290.

\appendix
\section{Power law mass function}\label{App.Power}

\begin{figure}[htb]
\includegraphics[width=0.9\columnwidth]{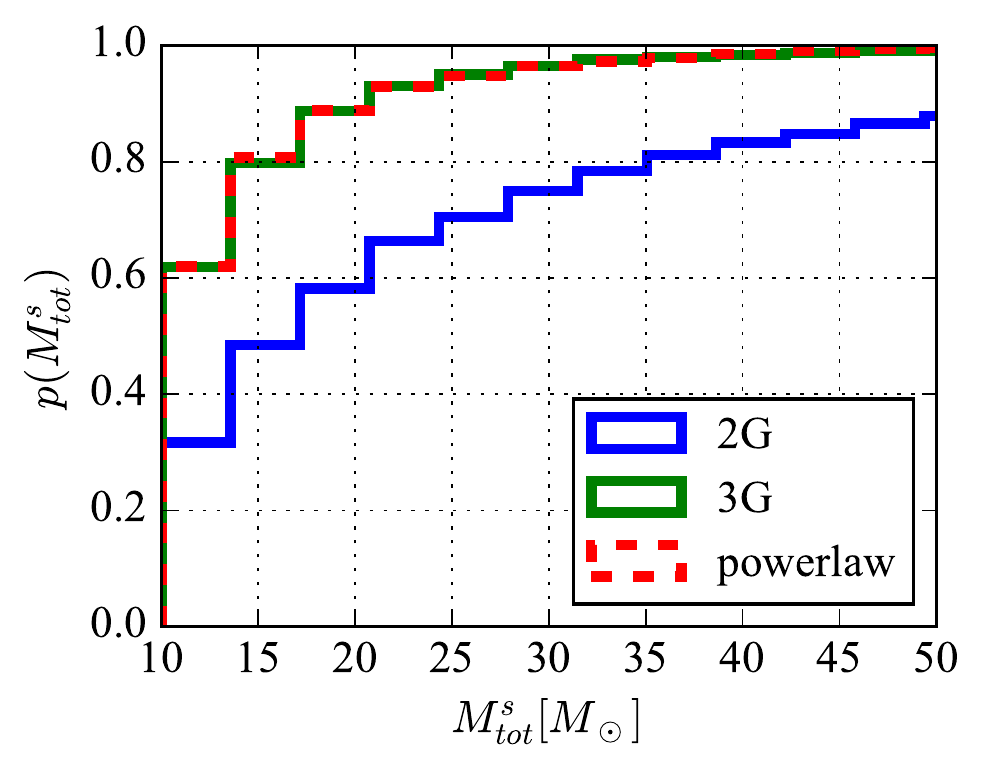}
\caption{The cumulative source-frame total mass distribution  of detectable events with a 2 interferometers network of advanced detectors at design (2G) or CE-like (3G). The underlying population is a power law with index 2.5 as described in the text, and shown in the plot as a dashed line.}
\label{Fig.AppMass}
\end{figure}

Throughout this paper we have considered a mass distribution uniform in both total mass and mass ratio. This was done in order no to introduce any selection effect from poorly constrained astrophysical quantities. In this appendix we wish to explore a particular alternative mass function, and show how all the results we found in the body of the paper still hold true, with minor differences.
We have thus considered the same mass distribution studied by the LIGO and Virgo collaborations in Ref.~\cite{O1BBH}. In particular, the distribution of the primary object is a power law with index $\alpha$: $p(m_1)\sim m_1^{-\alpha}$, while the distribution of the secondary object is uniform between 5~\msun{} and $m_1$.
A further constraint is imposed that $m_1+m2\leq100$~\msun. In Ref.~\cite{O1BBH} the median $\alpha$ and its the 90\% credible interval was estimated to be $\alpha=2.5^{+1.5}_{-1.6}$. We generated a population with $\alpha=2.5$. 
Naturally, we still find a that the 2G detector has a selection bias for heavier objects. In Fig.~\ref{Fig.AppMass} we show the cumulative distribution for the total mass of detected events by 2G and 3G, together with the underlying power law population. We see that 3G networks can detect the whole population without apparent bias (this could have been guessed from the fact that the left side of Fig.~\ref{Fig.SourceMass} was flat). On the other hand, 2G detectors will detect more heavy objects. While for 3G detectors 90\% of sources will have total mass below 20~\msun{} (exactly as in the true population), 2G detectors will only detect 60\% of events below that mass. 

\begin{figure}[htb]
\includegraphics[width=0.9\columnwidth]{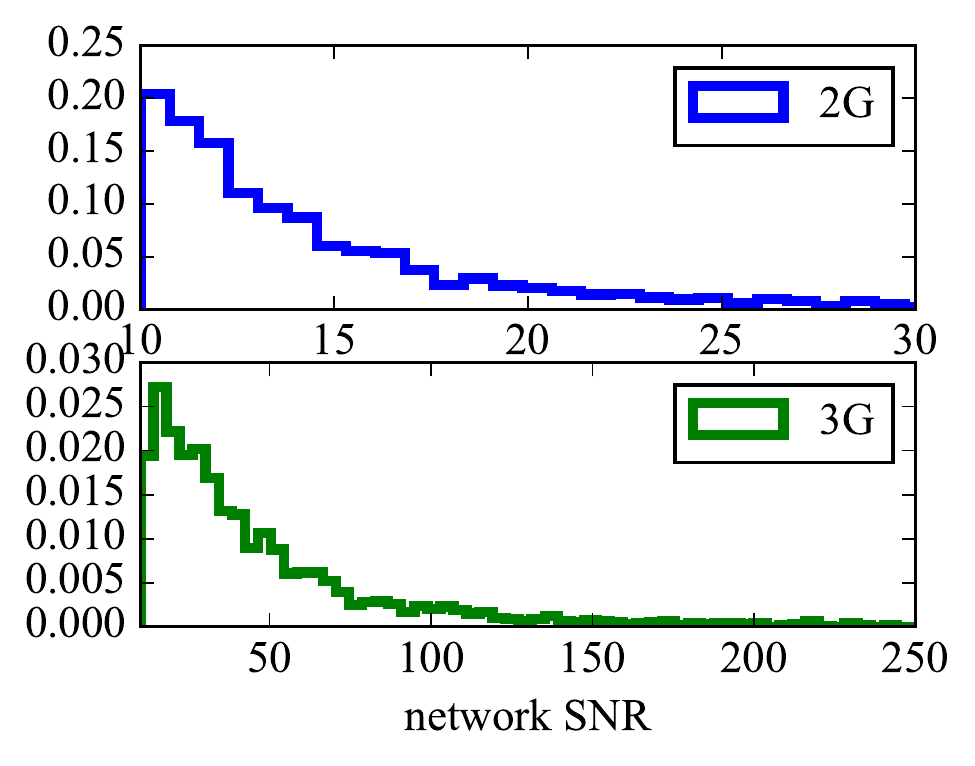}
\caption{The network SNR distribution for the events described in this appendix. }
\label{Fig.AppSNR}
\end{figure}

Even for the power law mass distribution, we find that the inclination angle distribution will look like in Fig.~\ref{Fig.Iota} since all events will be detectable.
Finally, we find that it is still the case that the peak of the SNR distribution for the 3G network will be at a value above threshold, although lower than if the mass distribution extended to IMBH, as in the main text. This is show in Fig.~\ref{Fig.AppSNR}, where a peak is visible at SNR of \si15 network. It also worth noticing that very large SNRs would be less frequent than what see in Fig.~\ref{Fig.SNRs}, due to the lack of heavy BBH sources.

\section{Binary formation rate}

In this study we have assumed the merger rate of BBH is flat in the cosmic history. While this is a simplified assumption, it does not impact at all the results in Sec.~\ref{Sec.Iota}, since these are only a consequence of the fact that 2G detectors do not reach the region of the universe where most sources are, while 3G do. This would not become false if a specific realistic merger rate were folded into the analysis.
For the same reason, the results in Sec.~\ref{Sec.SNR} will also stand.
The main impact of a non-flat merger rate would be to slightly change the exact amount of mass selection bias, without changing the fact that the ``direction" of the bias will be reversed in 2G and 3G. In fact, assuming a different mass function changes our original setup more dramatically than a different formation rate, but leads to the same qualitative results, Appendix~\ref{App.Power}.

\bibliography{pe.bib}

\end{document}